\documentclass[12pt]{iopart}
\usepackage{graphicx}
  \newcommand{\snn}{\sqrt{s_{NN}}}

\newcommand{\s}{\sqrt{s}}
\newcommand{\pp}{pp}
\newcommand{\epem}{e^+e^-}
\newcommand{\pbar}{\overline{p}}
\newcommand{\pbarp}{\pbar p}
\newcommand{\qbar}{\overline{q}}
\newcommand{\qbarq}{\qbar q}
\newcommand{\cbar}{\overline{c}}

\newcommand{\nch}{N_{ch}}
\newcommand{\np}{N_{part}}

\newcommand{\nc}{N_{coll}}

\newcommand{\jpsi}{J/\psi}
\newcommand{\kpi}{K^{+}/\pi^{+}}
\begin{document}

\title[Thoughts on Heavy Quark Production]{Thoughts on Heavy Quark Production}

\author{Peter Steinberg\dag\
\footnote[3]{Correspondence: peter.steinberg@bnl.gov}
}

\address{\dag\ Chemistry Department, Brookhaven National Laboratory, Upton, NY 11973}

\begin{abstract}
Various aspects of heavy flavor production
in heavy ion collisions in the context of elementary collisions are reviewed.
The interplay between theory in experiment in $\epem$ and $\pbarp$ data is
been found to be non-trivial, even with new NLO calculations.
Quarkonium suppression in p+A and A+A show puzzling features, apparently
connected to features of inclusive particle production.
Open charm is found to scale as expected for a hard process, but
strangeness is also found to share these features.  These features
contribute to heavy flavor as a interesting probe of strong interactions.

\end{abstract}




\section{Introduction}


To understand the broad interest in non-light flavors in hadronic
collisions, one only has to appreciate two basic facts:
1) strangeness and charm are degrees of freedom which are not 
present in the initial colliding systems (predominantly composed
of up and down quarks) and 2) the successively larger quark masses
require larger momentum transfers
in their production processes, thus making the use of pQCD techniques
more reasonable as the strong coupling constant decreases.
These features make heavy flavor an interesting probe of the early
stages of heavy ion collisions, possibly sensitive to the features of
the produced, strongly-interacting medium.

\section{Charm Production in Elementary Collisions}

To even begin to address the issue of heavy flavor production in
heavy ion collisions, one must understand the essential features
of the production process in elementary collisions, including
$\epem$, $\pbarp$, and $e+p$ reactions.  Since the discovery of
the $\jpsi$ (closed charm) in 1974~\cite{Aubert:1974js,Augustin:1974xw}
and the subsequent discovery 
of D mesons (open charm) a year later, a large data set on charmed hadrons
has been built up over a variety of collision energies and reactions.
Only a small subset relevant to later arguments will be discussed
here.


\subsection{$\epem$ Collisions}
The apparently simple reaction of electron-positron annihilating
to a virtual photon and subsequently into hadrons ($\epem \rightarrow
hadrons$) is an excellent laboratory to study both the production
of heavy quarks from the QCD vacuum as well as the cleanest
environment to study their fragmentation into hadrons.
It provided the cleanest environment to discover the
$\jpsi$ and to measure its properties, as well as the
means to produce the higher-mass and higher-spin states of the
``charmonium'' spectrum, whose striking similarity to the 
already-known positronium level diagram was one of the early
great successes of the quark model of hadrons~\cite{perkins}.

The fragmentation of heavy quarks is an active field of research
in recent years.  This is a consequence of the recent turn-on of
B-factories at KEK and SLAC, as well as recent analyses
of production at the $Z_0$ pole with extensive particle identification.
One striking feature of heavy quark fragmentation which distinguishes
it from light-quark fragmentation the distinctive ``hard'' fragmentation
functions, with the leading particle taking typically 60\% of the
initial quark energy for charm (at B-factory energies), 
and nearly 80\% for beauty (at LEP energies)~\cite{Biebel:2002es}.  When
studied as a function of particle species in b and c jets, 
one finds a striking example of the ``leading particle effect'', 
where kaons appear to carry the memory of the initial heavy quark
(clearly by virtue of the cascading weak decays of the heavy quarks
$b \rightarrow c \rightarrow s$)~\cite{Abe:2003iy}. 
One also sees a dramatic depletion
of all particles in the forward direction, a direct indication of
the so-called ``dead-cone'' effect~\cite{Dokshitzer:1991fd}.  Both of these
effects (hard fragmentation and dead-cone) will be relevant for later
discussions of heavy quark production in heavy ion collisions.

\begin{figure}[t]
\begin{center}
\begin{minipage}{6cm}
\includegraphics[width=6cm]{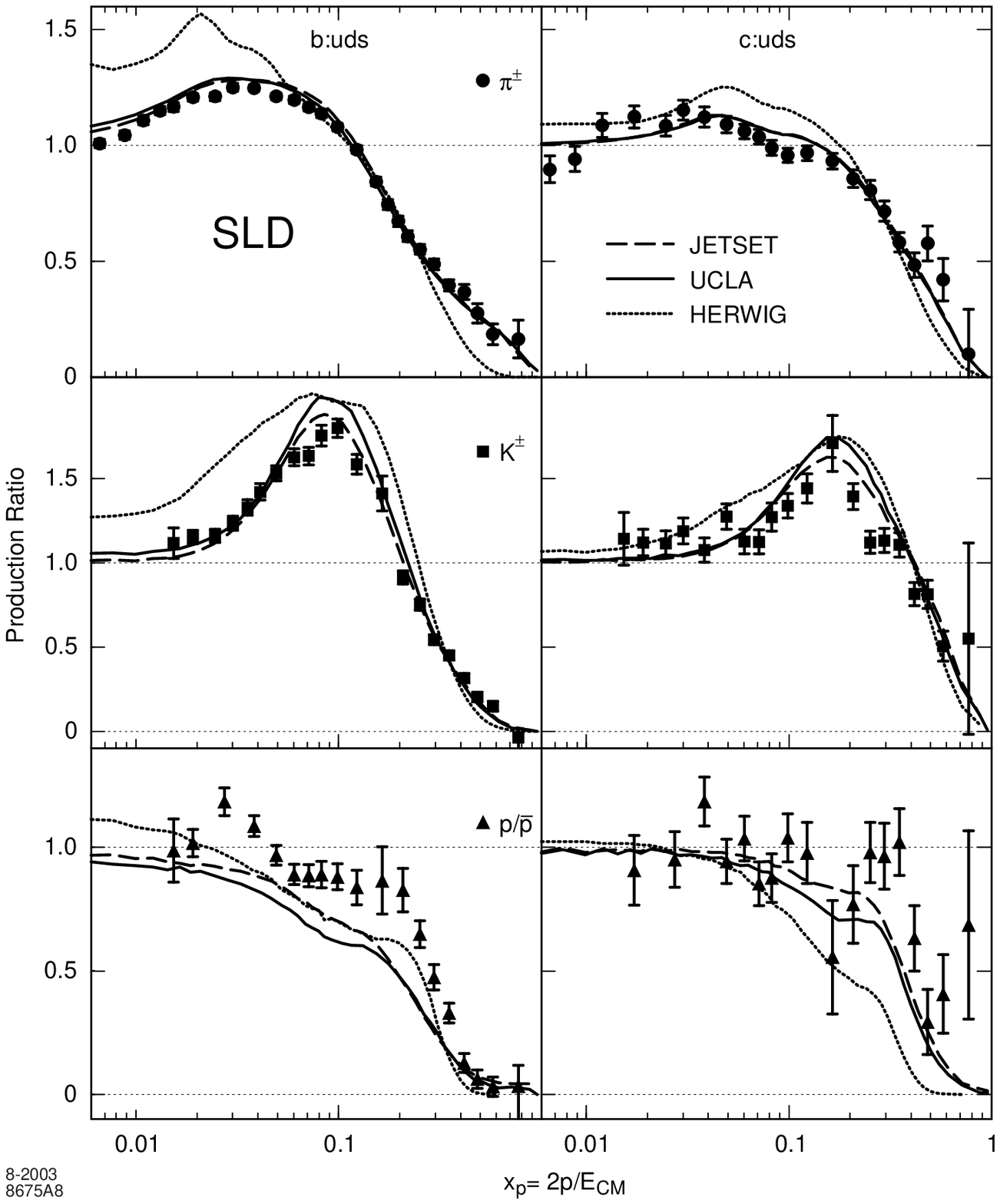}
\caption{
Identified-particle fragmentation functions for $b$ and $c$ quarks measured
by the SLD experiment.
\label{pub9949-fig09}
}
\end{minipage}
\hfill
\begin{minipage}{9cm}
\includegraphics[width=9cm]{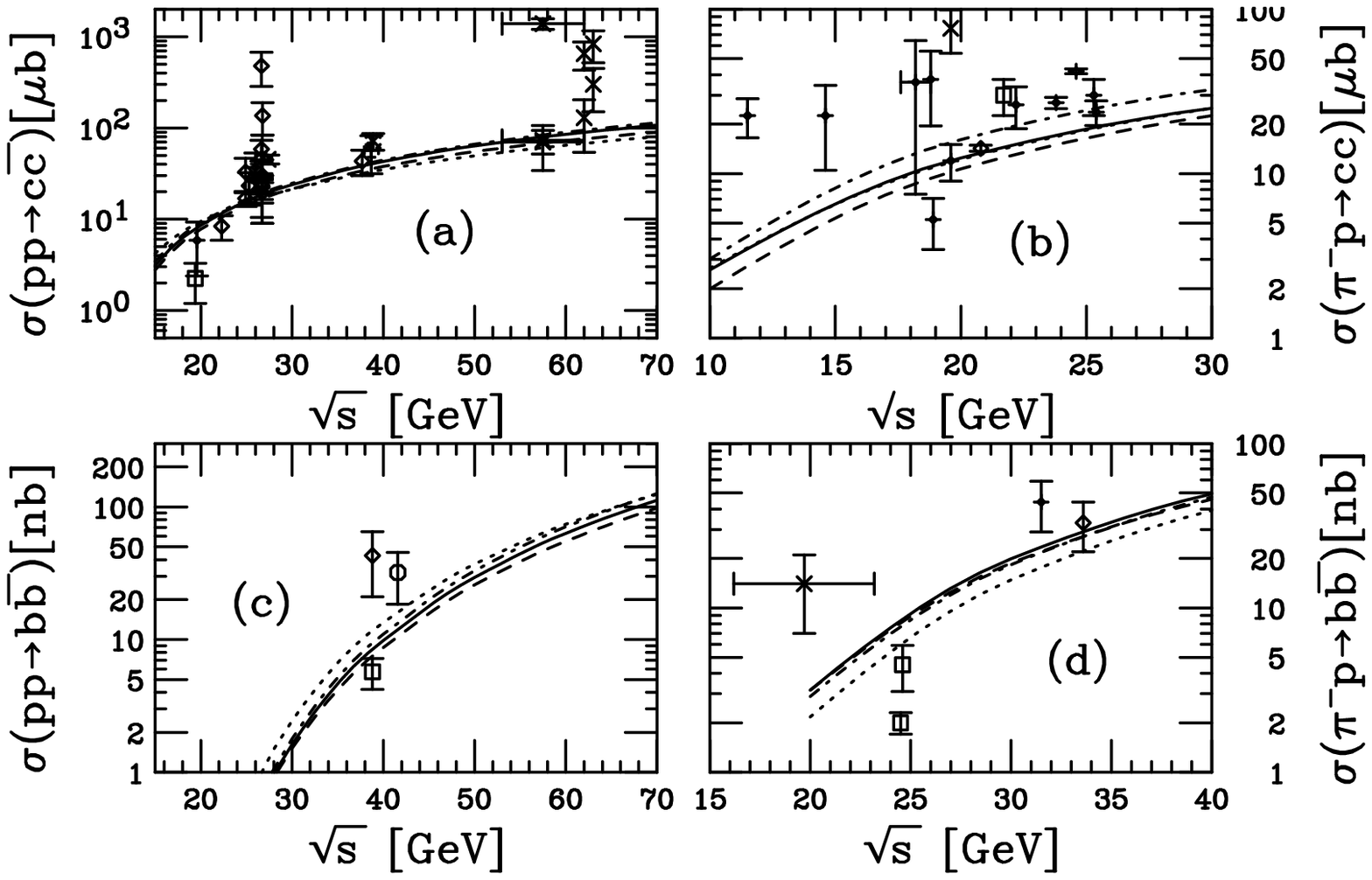}
\caption{
pQCD calculations for charm and bottom production compared with
experimental data at fixed target energies.
\label{vogt-ccbb}
}
\end{minipage}
\end{center}
\end{figure}

As a preparation for future discussions of $\jpsi$ production, it is 
important to discuss the level of our understanding of how
such particles are formed from heavy quarks.  Originally it was
thought that a charm and anti-charm quark were produced simultaneously
via $qq$ or $gg$ scattering and then the pair would ``coalesce'' into
a color singlet $\jpsi$ - the ``color singlet model'' (CSM)~\cite{Bodwin:1994jh,color-singlet}.  
However, it turned out that this model
in conjunction with pQCD production estimates dramatically under-predicted
the rates in $\pbarp$ collisions (to be discussed below).  This led to
the ``color octet model'' (COM)~\cite{color-octet}
which attributed $\jpsi$ production to
the correlated production of a $\cbar c$ precursor state already
close in phase space, color neutralized by an available soft gluon.  
By reducing the difficulty
of situating the produced quarks near enough in phase space, the COM predicted
much larger production rates. A similar model is the ``color evaporation
model'' (CEM)~\cite{color-evaporation}, 
which predicts that a $\jpsi$ is dominantly produced via
gluon fragmentation, presuming that one can always find a soft gluon.
This model also predicts simple formulas for $\sigma(\jpsi)/\sigma(c\cbar)$.

In this context, one can appreciate the
puzzle in $\jpsi$ production found recently in 
BELLE data~\cite{Abe:2002rb}.  
They found a striking correlation in the production
of $\jpsi$ with other charmed particles.  The fraction 
$R=\sigma (\epem \rightarrow \jpsi + c\overline{c})/
\sigma (\epem \rightarrow \jpsi + X)=0.59^{+0.15}_{-0.13}\pm0.12$
is a striking challenge for both singlet and octet models of $\jpsi$
production.  Neither would predict that a charm pair, which should
normally hadronize into jets with leading D mesons, would prefer to
emit {\it another} charm pair.  However, it is the opinion of this
author (perhaps unoriginally) that this large ratio may simply be
a sort of ``trigger bias''.  Since a decay of a virtual photon
into oppositely pointing $c$ and $\cbar$ jets would preclude these
quarks from creating a bound state, it seems unlikely to ever
produce a $\jpsi$ without the liberation of yet another pair
of quarks.  In this context, it may be difficult to produce
a $\jpsi$ without additional charm production, but this heuristic
argument (which is ultimately based on quark counting) 
does not have any substantial theoretical underpinning at present.

\subsection{Charm Production in p+p collisions}

In $p+p$ collisions, the dominant charm production process is the
scattering of quarks and gluons in the initial-state parton
structure of the projectiles.
Charm production in $\pp$ and $p+A$ collisions thus offers a means to
explore a large range of kinematic variables ($x$ and $Q^2$) at
the cost of needing to introduce substantial phenomenological
input to interpret the data.  That is, at the same time as one
is testing the utility of the fundamental pQCD cross sections,
one is also testing our understanding of non-perturbative
structure and fragmentation functions.

As the most basic example of the concept of ``collinear factorization'' ,
one can examine a typical expression of the total charm cross
section in its factorized form (see e.g. Ref.~\cite{Bedjidian:2003gd}:

\begin{eqnarray}
\sigma(S,m^2_Q) & = & 
K \times \sum_{i,j=q,\qbar,g}\int^{1}_{4m^2_Q / s} \frac{d\tau}{\tau}
\int dx_1 dx_2 \delta(x_1 x_2 - \tau)\\
\nonumber & \times & f^A_i (x_1, \mu^2_F) f^B_i (x_2, \mu^2_F)
\times \sigma_{ij} (s,m^2_Q, \mu^2_F, \mu^2_R).
\label{blah}
\end{eqnarray}

This formula shows how one factorizes out the initial state
of the projectiles from the fundamental pQCD production
cross section, constraining the kinematics with the delta
function.  The ``projectile'' and ``target'' are incorporated
separately via their own structure functions and the respective
momentum fractions sampled from each ($x_1$ and $x_2$).
The ``K-factor'' is required to match the overall normalization,
which is often underestimated by fixed-order calculations, but
is expected to be unnecessary if all orders could be included.
Results from these calculations are shown for p+p collisions
in Fig.~\ref{vogt-ccbb}.

To extract single-particle spectra, one must convolute the
terms which describe the production of charm quarks with
the fragmentation functions we have already discussed in the
context of $\epem$ annihilations to charm or bottom quarks.

\begin{eqnarray}
d\sigma & \propto & 
K \times f^A_i (x_1, \mu^2_F) f^B_i (x_2, \mu^2_F) \times 
d\sigma_{ij} (s, m^2_Q, \mu^2_F, \mu^2_R)\\
\nonumber & \nonumber & \times D(z,\mu^2) \times g(k^2_T)
\label{blah2}
\end{eqnarray}

Here we see how particle spectra result from 3 independent steps:
1) production controlled by the initial nucleon or nuclear
structure, 2) pQCD cross sections, 3) fragmentation of the
produced partons.  It should be noted that the scale dependence
of the cross sections and fragmentation functions are themselves
typically controlled by DGLAP QCD evolution, which provides for
additional low-x gluons as the energy increases.  This breaks
the initially-proposed Feynman scaling and leads to the 
production of very soft gluons near $x=0$.  It should also be
noted that experimental transverse momentum spectra are much softer
than the theoretical predictions.  To account for this, 
one typically adds a phenomenological ``$k_T$-broadening'',
shown as $g(k^2_T)$, which harden the spectrum and increase
the cross section.  The same does not appear to be needed
in the longitudinal distributions at fixed target energies,
which is a real puzzle at present~\cite{vogt}.

It should be emphasized in this context that QCD ``factorization'',
in the formal sense established by Collins, Soper, and Sterman
\cite{Collins:1985ue},
is critical concept for the application of pQCD to hadronic and
heavy ion collisions.  It is only officially a well-defined
concept for very large momentum transfers, such that 
$Q^2 \gg \Lambda^2_{QCD}$.  In this regime, the momentum
transfers take place over small space-time distances and
one may prove that in certain conditions (e.g. Drell-Yan
production) that the ``hard'' pQCD cross sections are
independent of the long-wavelength fluctuations in the
proton wave function.  This is thus a prerequisite for
any kind of nucleon or nuclear structure physics, although 
it is also critical that the structure functions are 
``universal'' and independent of which pQCD subprocess
one is considering.  Finally, if all this is true, then 
hard processes in $p+A$ and $A+A$ collisions become comprehensible as a 
superposition of the independent binary collisions given
the initial impact parameter.

\subsection{$\pbarp$ Collisions}

\begin{figure}[t]
\begin{center}
\includegraphics[width=12cm]{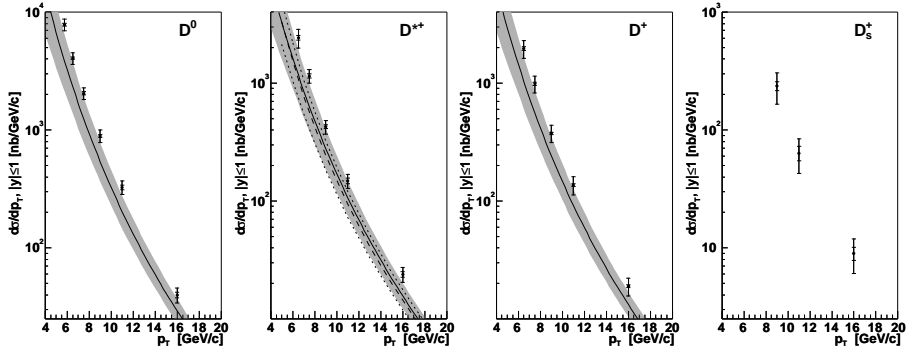}
\caption{
Cross sections for D mesons, measured by the CDF experiment
and compared with NLO pQCD calculations.
\label{DiffXsection_combined_4x1}
}
\end{center}
\end{figure}

It is generally understood, by means of the QCD factorization
theorems, that hadron-hadron interactions at asymptotically
high energies should be amenable to a perturbative QCD
description, especially for processes involving large 
momentum transfers.  
Historically, this expectation has been well established for
high-$E_T$ jet rates but typically confounded for
open heavy flavor and quarkonium production
at the highest-available energies at the Fermilab Tevatron
($\s=1.8$ TeV).

Open charm at CDF has only recently come under reasonable
theoretical control.  The data shown in Fig.~\ref{DiffXsection_combined_4x1}
on D meson production~\cite{Abe:2002rb} is compared with NLO calculations,
and one finds that the data sit on the upper edge of the theoretical
systematic error bands (shown in gray).  These are the errors
generated by varying the factorization and renormalization scales
($\mu_F$ and $\mu_R$) up and down by a factor of two.  Thus, one
is finding a consistency between data and theory of about
50-100\%, which is certainly not in the realm of precision physics
yet, but NLO calculations of heavy flavor production are maturing
rapidly, so more improvements should be expected.

The situation with $\jpsi$ production has been steadily improving
since the realization in the early 1990's that pQCD calculations joined
with the CSM under-predicted the measured $\jpsi$ production rates at
very high $p_T$ (6-20 GeV/c) by enormous factors (1-2 orders of
magnitude)\cite{Abe:1997jz}.  
The introduction of the COM improved the agreement 
in rates, but predicted a polarization which has been strongly
contradicted by recent Tevatron data\cite{Acosta:2003ax}.

The main point to draw from these
discussions is that even in the highest energy elementary collisions, 
heavy flavor production is not trivial.  
This situation should be kept in mind
when applying similar calculations to lower-energy heavy ion collisions,
where the momentum transfers are smaller, and the collision systems are
substantially more complicated.

\section{Charm Production in p+A and A+A Collisions}

\begin{figure}[t]
\begin{center}
\includegraphics[width=9cm]{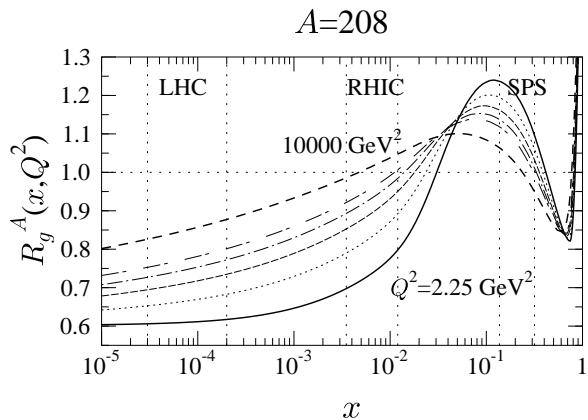}
\caption{
Nuclear shadowing calculations from Ref.~\cite{Eskola:2001gt}, showing
the $Q^2$ evolution.
\label{Rgdist}
}
\end{center}
\end{figure}

Given the current understanding of heavy flavor
production in high-energy collisions, it would seem
difficult to get a handle on the nature of similar production
processes in the more complicated environment of a heavy ion collision.
To this end, studies of proton collisions on nuclei are used to 
get a potentially more relevant baseline for A+A collisions.

\subsection{Nuclear Shadowing}

One of the complications of studying flavor production in a nuclear
environment is the nucleus itself.  Deep-inelastic experiments in the
early 1980's revealed that the effective structure of a nucleon
embedded in a nucleus (expressed as ratios such as $F_2^A/F_2^d$) were
substantially modified relative to the nucleon structure.  
While only a modest dependence on $Q^2$, the hardness scale of the 
scattering process, was observed\cite{Adams:1995is}, 
a dramatic dependence was found
in the $x$ variable, representing the momentum fraction of the
partons struck by the incoming virtual photon.  At the highest
$x$ range ($x>0.9$), 
one observes a large enhancement due to Fermi smearing
of the nucleons in the nucleus.  
At moderate $x$ ($0.2<x<0.9)$ one finds a depletion of the parton
distributions, a phenomenon called the ``EMC effect''\cite{Aubert:1983xm}
but which continues to resist 
simple explanation even now.  To maintain energy-momentum
conservation, the strong depletion in moderate $x$ partons must be
compensated by a rise in $x$ values of approximately $0.1$ ,called
the ``enhancement'' or ``anti-shadowing'' region.
Finally, the region of very low $x$ ($x<0.1$) shows
``shadowing'' phenomenon, characterized by a depletion nearly
flat down to very low $x$ values.

While no unique explanation of this
effect exists, it is generally thought to arise by the quantum-mechanical 
interference of the wave functions of the nucleons
in the nucleus, which make up the tube in front of the incoming
virtual photon (or equivalently $\qbarq$ state)\cite{Mueller:1985wy,
McLerran:2003yx}.

By treating nuclear effects as modifications of the initial state,
rather than as a dynamical effect of the collision of the probe
with the nucleus, the physics of nuclear modifications is generally
considered to not
be in the realm of perturbative QCD.  Thus, ``nuclear shadowing'' a
term which encompasses all of the above effects and denotes any
modification of the structure function of a nucleon in a nucleus,
is generally handled by a judicious parametrization of existing
data, using known pQCD techniques to evolve the modification factors
as a function of $x$ and $Q^2$.  This is the basis of the
well-known EKS98 approach.  They take the modification factors
$R^A_i$ for parton $i$ and evolve using DGLAP techniques\cite{Eskola:1998df}.
These lead to a set of curves shown in Fig.~\ref{Rgdist}, from Ref.
\cite{Eskola:2001gt}, which
clearly show the ``shadowing'', ``anti-shadowing'', ``EMC effect''
and ``Fermi motion'' regimes, but now as a function of $Q^2$
from $2.25 GeV^2$ to $10000 GeV^2$.
It should not be forgotten, however, that saturation models do offer
a QCD-based model for shadowing phenomenon with some basis
in perturbative QCD.

\subsection{Charm as a ``Hard Probe''}

As we showed in above sections, heavy flavor is not fully under
control even in elementary reactions.  Open charm is still under-predicted 
by pQCD calculations, requiring substantial
$K$ factors to agree with data.  Even then, special fragmentation
functions and intrinsic $k_T$ are needed to describe the differential
cross sections.  Closed charm is plagued by its
reliance on NRQCD models of $\jpsi$ formation.  How could we then 
expect to use it as a ``calibrated'' probe in a heavy ion collision?
The answer again relies on QCD factorization: in principle the short-wavelength
field configurations reflected in large transverse
momentum processes are
only moderately sensitive to the longer-range configurations typical
of soft processes.  This leads to the prediction that high-$p_T$
phenomena, including charm production rates, are only sensitive to 
the number of binary collisions experienced by the nucleons as the
nuclei interpenetrate each other.  To be more specific, one would
expect the rate of these rare processes to scale {\it linearly} 
with the number of binary collisions, essentially measuring the
longitudinal thickness of the projectile and target nuclei seen
by the incoming nucleons.
Given this, strong deviations from factorization expectations can then be
interpreted a indications of non-trivial nuclear physics.  
$\jpsi$ suppression is the canonical example of this story\cite{Abreu:2000ni}.

Just as a reminder, we briefly review the two dominant scaling
variables for yields in heavy ion collisions.  Soft particle 
production in $p+A$ and $A+A$ collisions has 
been observed to scale linearly with the number
of ``participating'' nucleons ($\np$), i.e. nucleons
which undergo any inelastic process.  Hard processes, by contrast,
are expected to scale with the number of ``binary collisions'',
as mentioned in the previous paragraph.  These two quantities are
tightly correlated by the fact that at smaller impact parameter
($b$), a typical nucleon sees a larger thickness, 
so the larger the overlap volume, the greater the thickness
of nuclear matter it collides with, giving an approximate number
of collisions that scales as volume$\times$thickness $\sim
R^4 \sim \np^{4/3}$.  Glauber calculations show that 
over the range of impact parameters
measured by RHIC experiments, one is sampling thicknesses of
$\nu = \nc/(\np/2) \sim 1-6$, as shown in Fig.~\ref{glauber_200_ncnp_npnc}.

\subsection{$\jpsi$ Suppression}

\begin{figure}[t]
\begin{center}
\begin{minipage}{9cm}
\includegraphics[width=9cm]{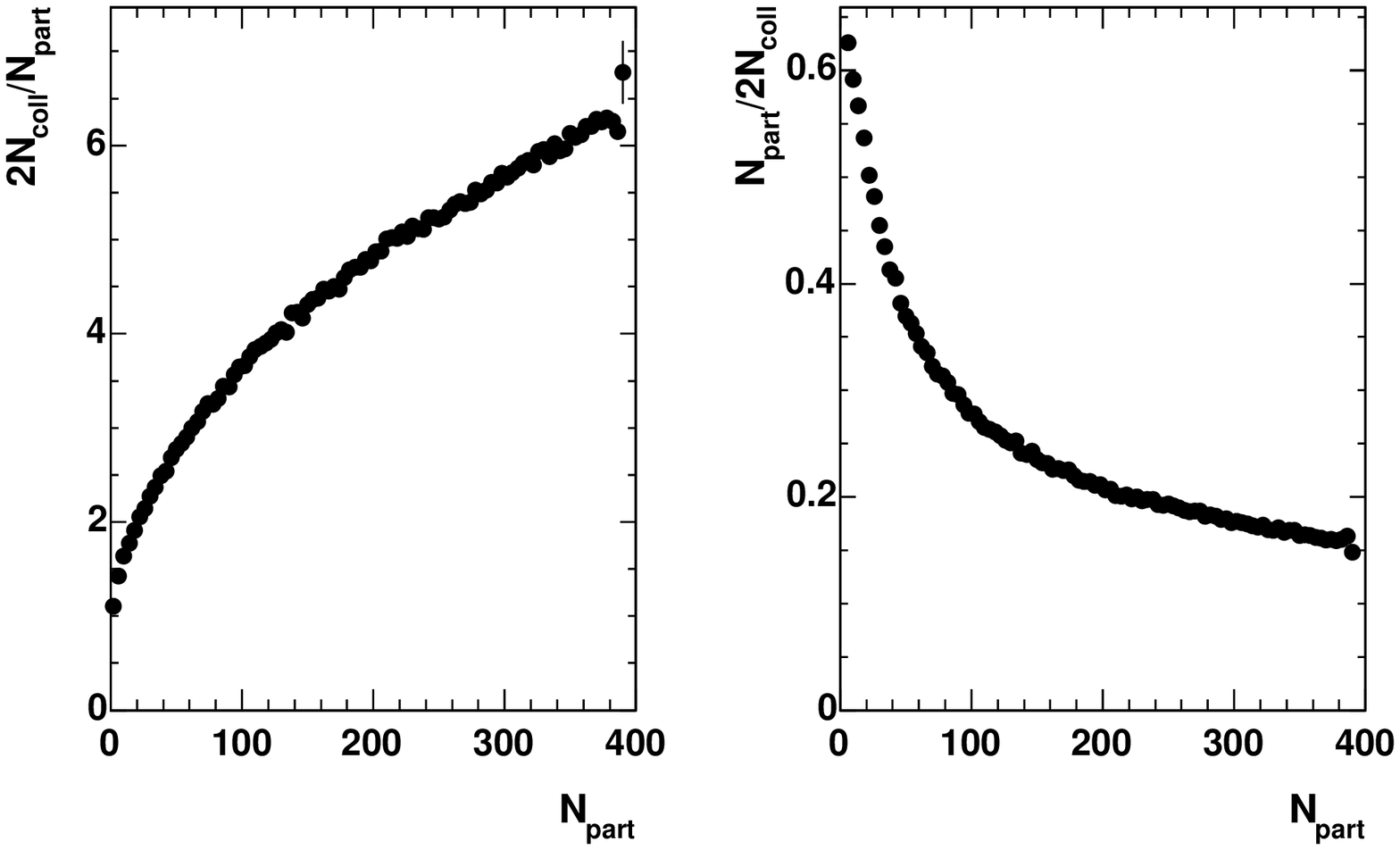}
\caption{
Glauber model calculations showing a) the number of collisions per
participants, and b) the number of participants per collision.
\label{glauber_200_ncnp_npnc}}
\end{minipage}
\hspace{\fill}
\begin{minipage}{6cm}
\includegraphics[width=6cm]{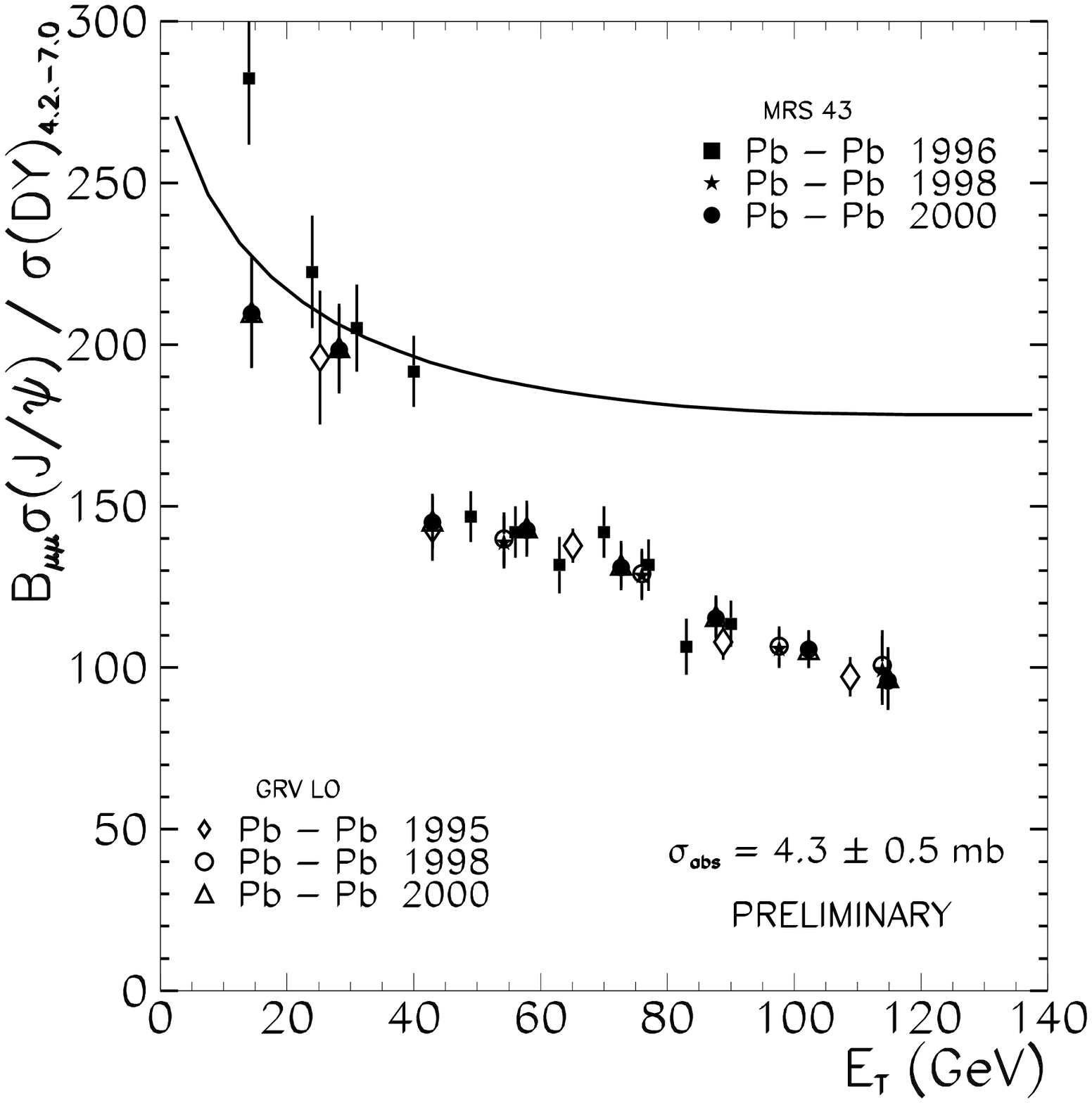}
\caption{
NA50 results on J/Psi suppression, showing the ratio of $\jpsi$
production Drell-Yan as a function of centrality.
\label{tmundo_preto}
}
\end{minipage}
\end{center}
\end{figure}

The NA50 analysis of $\jpsi$ suppression is generally shown as
the ratio of $\jpsi$ production relative to Drell-Yan production
a function of a centrality variable (typically $E_T$)\cite{Abreu:2000ni}.  
This ratio is then compared with a calculation of the expected
``normal nuclear suppression'' extracted from p+A data, as shown
in Fig.~\ref{tmundo_preto}.
The data appear to agree with the normal suppression in peripheral
events, but then deviate strongly for moderately peripheral events
saturating at a suppression factor of approximately 50\%.
These results are insensitive to various choices of the centrality
variable (i.e. a rescaling of the horizontal axis)~\cite{Colla},
which is not surprising considering that these ratios
are made within each centrality class.

While these results have been cited widely and studied by a variety
of theoretical approaches, it makes sense to look very carefully
at the systematics of $\jpsi$ production from other systems and
energies, by which ``normal'' nuclear suppression is determined.
The analysis of this suppression is based on the
fact that the yield of $\jpsi$ per target nucleon tends to systematically
decrease as a function of nuclear thickness.  This is typically
interpreted as the absorption of produced $\jpsi$ in the nucleus 
itself, presumably by $\jpsi+N$ inelastic scattering~\cite{Kharzeev:1996yx}.
This absorption is described empirically by the parameter $\alpha$,
which is made from a fit of the yields to the form
$\sigma^{p+A}_{\jpsi} = \sigma^{0}_{\jpsi} \times A^\alpha$,
with $\sigma_0$ ideally the $p+p$ cross section (but not
necessarily so).
If $\alpha=1$ then every nucleon in the nucleus is visible to
the incoming proton, which is equivalent to $\nc$-scaling.
We interpret $\alpha<1$ as due to the shadowing
of the interior by the nucleons on the surface, i.e. $\alpha \sim 2/3$
corresponds to complete surface absorption, with even lower values
interpreted as further absorption in the bulk.

NA50 extracted values of $\alpha_{\jpsi} = 0.931\pm0.002\pm0.007$
\cite{Colla}.
implying a substantial absorption in heavy targets.
They also analyze these results in terms of an attenuation in
the effective thickness  of the nuclear target
(which is approximately $\rho LA^{1/3}$), $N\propto exp(-\sigma \rho L)$ and
using Glauber geometry as a function of centrality 
to determine $\rho L$ in order to estimate
$\sigma_{\jpsi+N}$.  This cross section was as large as $7.3\pm0.6$ mb in
the 1996 analysis, but has decreased to $4.3\pm0.6$ mb in
a recent analysis incorporating the S+U data\cite{Colla}.  
Thus, this cross section
appears to depend strongly on the fit assumptions, lending some doubt
to its status as representing a primordial physics process.

\begin{figure}[t]
\begin{center}
\begin{minipage}{75mm}
\begin{center}
\includegraphics[height=7cm]{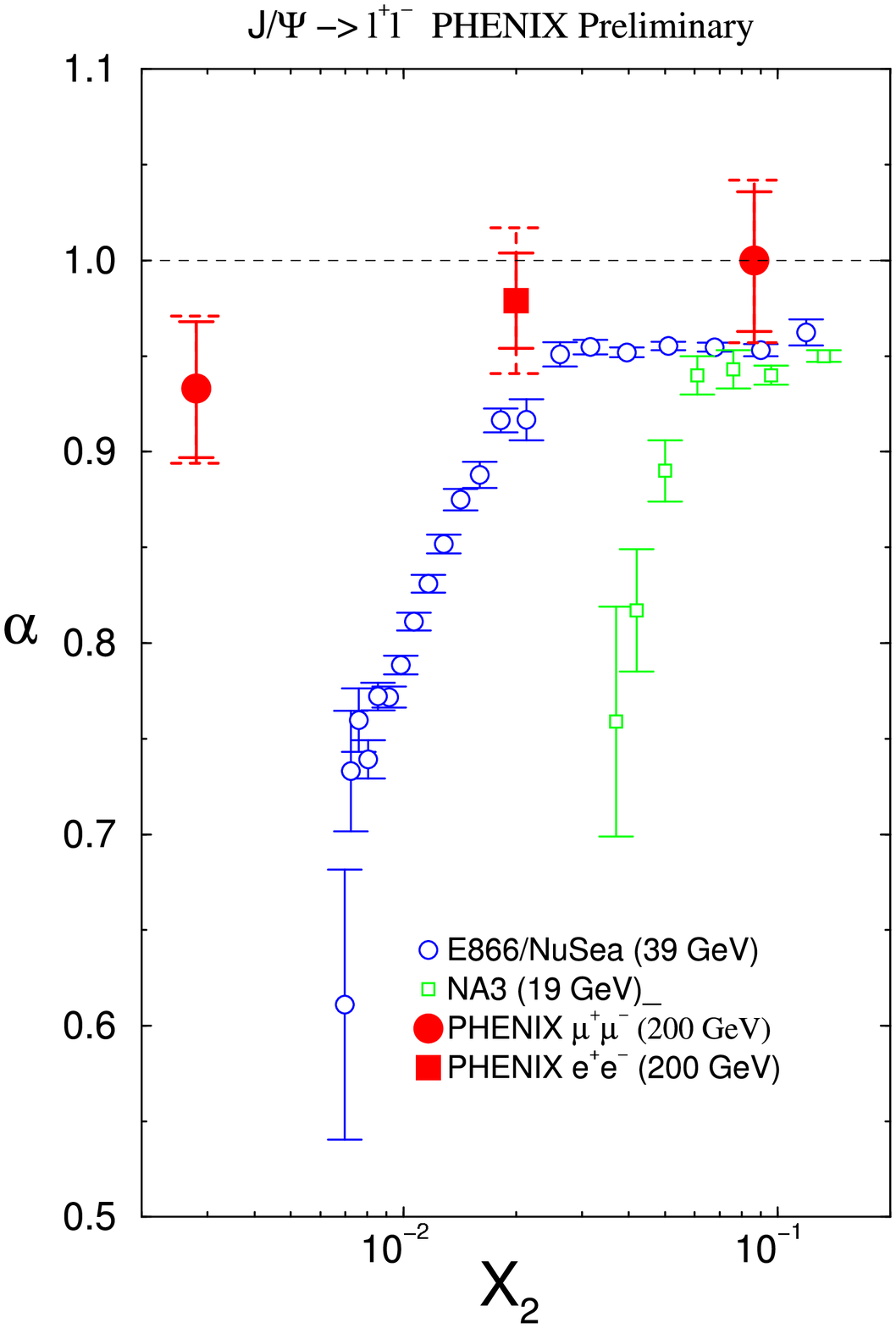}
\caption{
$\alpha$ as a function of $x_2$ for a variety of beam energies.
\label{alpha_x2}
}
\end{center}
\end{minipage}
\hspace{\fill}
\begin{minipage}{75mm}
\begin{center}
\includegraphics[height=7cm]{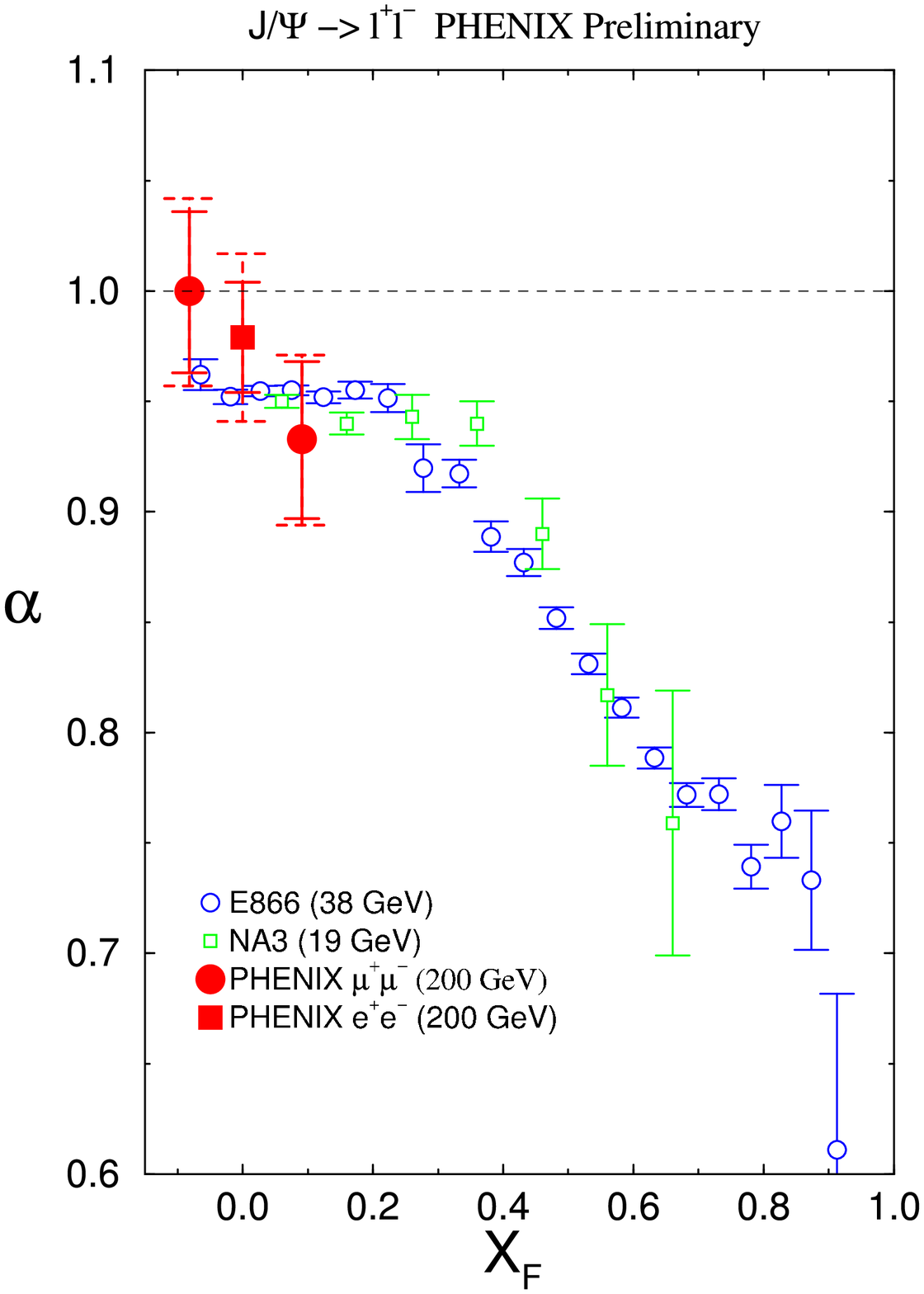}
\caption{
The same data for $\alpha$ as a function of $x_F$.
\label{alpha_xf}
}
\end{center}
\end{minipage}
\end{center}
\end{figure}

Some insight on this can be gained by study of other important
$\jpsi$ data sets from PHENIX at RHIC and E866 at Fermilab.
These experiments have larger kinematic acceptance than NA50,
E866 with the full forward range $y=0-4$ ($x_F>0$) in p+A collisions 
and PHENIX with forward and backward coverage $-2<y<2$ in d+Au collisions,
in order to elucidate the role of shadowing and energy loss in $\jpsi$
production.
If the depletion of
$\jpsi$ as a function of nuclear thickness was simply due to
the shadowing of the parton densities in the nuclear target,
then the suppression factor would be simply due to the nuclear properties
and thus $\alpha$ would be a universal function of $x_2$, 
the momentum fraction of the target parton, 
invariant with beam energy.  
As shown in Fig.~\ref{alpha_x2}, this expectation is dramatically violated
when one compares NA3, E866 and PHENIX (something already
noted previously by E866~\cite{Leitch:1999ea} comparing to NA3 data from
CERN).

This violation of $x_2$ scaling should be contrasted
with scaling which {\it is} observed in the forward region
between PHENIX, E866 and NA3 over a broad range in $x_F$, 
as shown in Fig.~\ref{alpha_xf}~\cite{deCassagnac:2004kb}.
While this would seem to be merely fortuitous, it should
not be forgotten that inclusive charged particle production
also seems to scale with $x_F$ across a wide range of beam
energies.  This is seen in the phenomenon of ``limiting 
fragmentation'', where particle yields seem to scale when
plotted as a function of $y^{\prime}=y-y_{beam}$~\cite{Back:2002wb}.
Thus, it seems that $\jpsi$'s also obey a sort of limiting fragmentation,
similar to soft processes, despite their presumed status as
a hard probe.

\begin{figure}[t]
\begin{center}
\begin{minipage}{90mm}
\includegraphics[width=90mm]{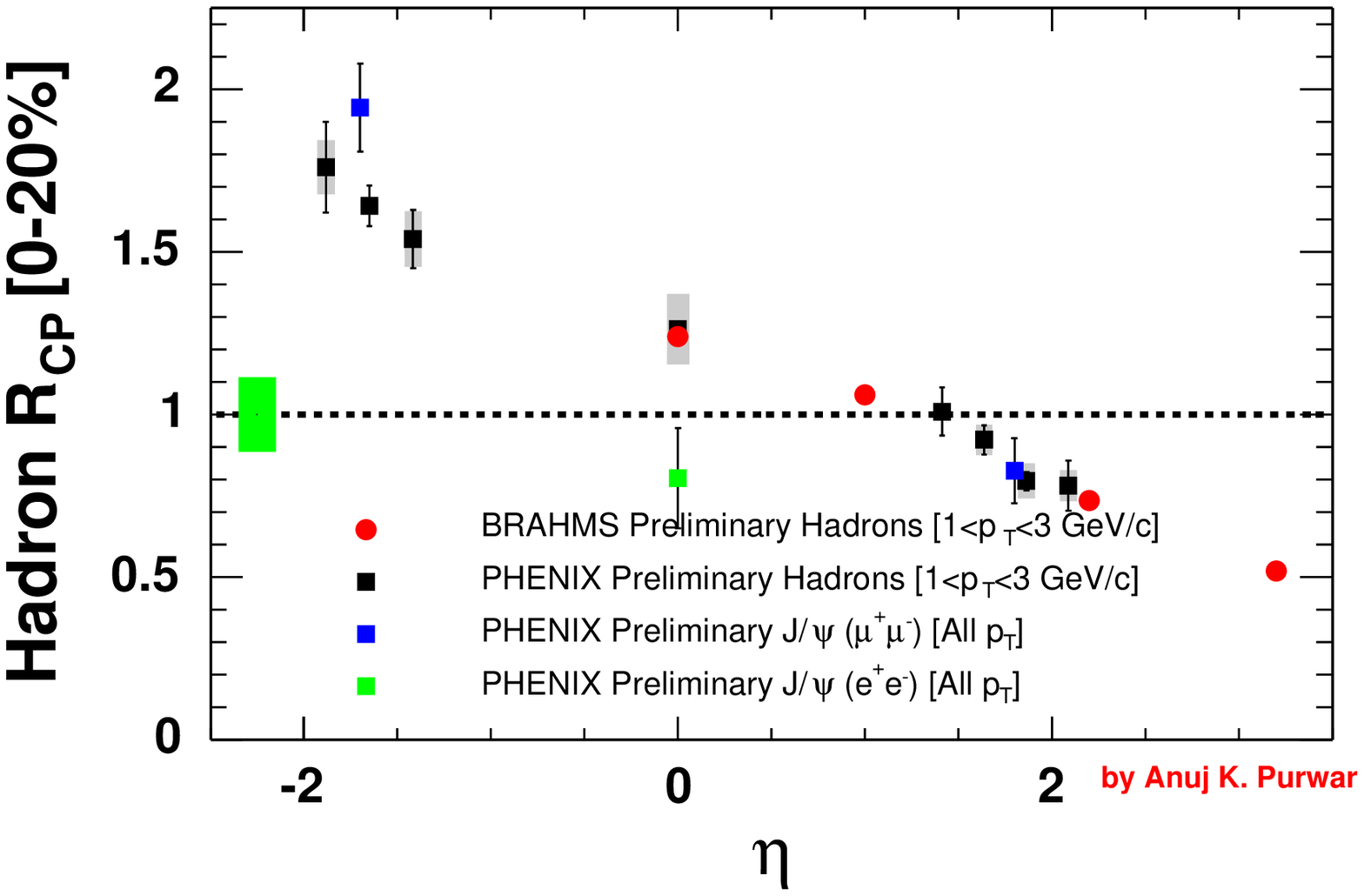}
\caption{
$R_{CP}$ measured in 200 GeV d+Au collisions at RHIC, both
for charged hadrons and $\jpsi$.
\label{brahms_phenix_rcpJpsi}
}
\end{minipage}
\hspace{\fill}
\begin{minipage}{60mm}
\includegraphics[width=60mm]{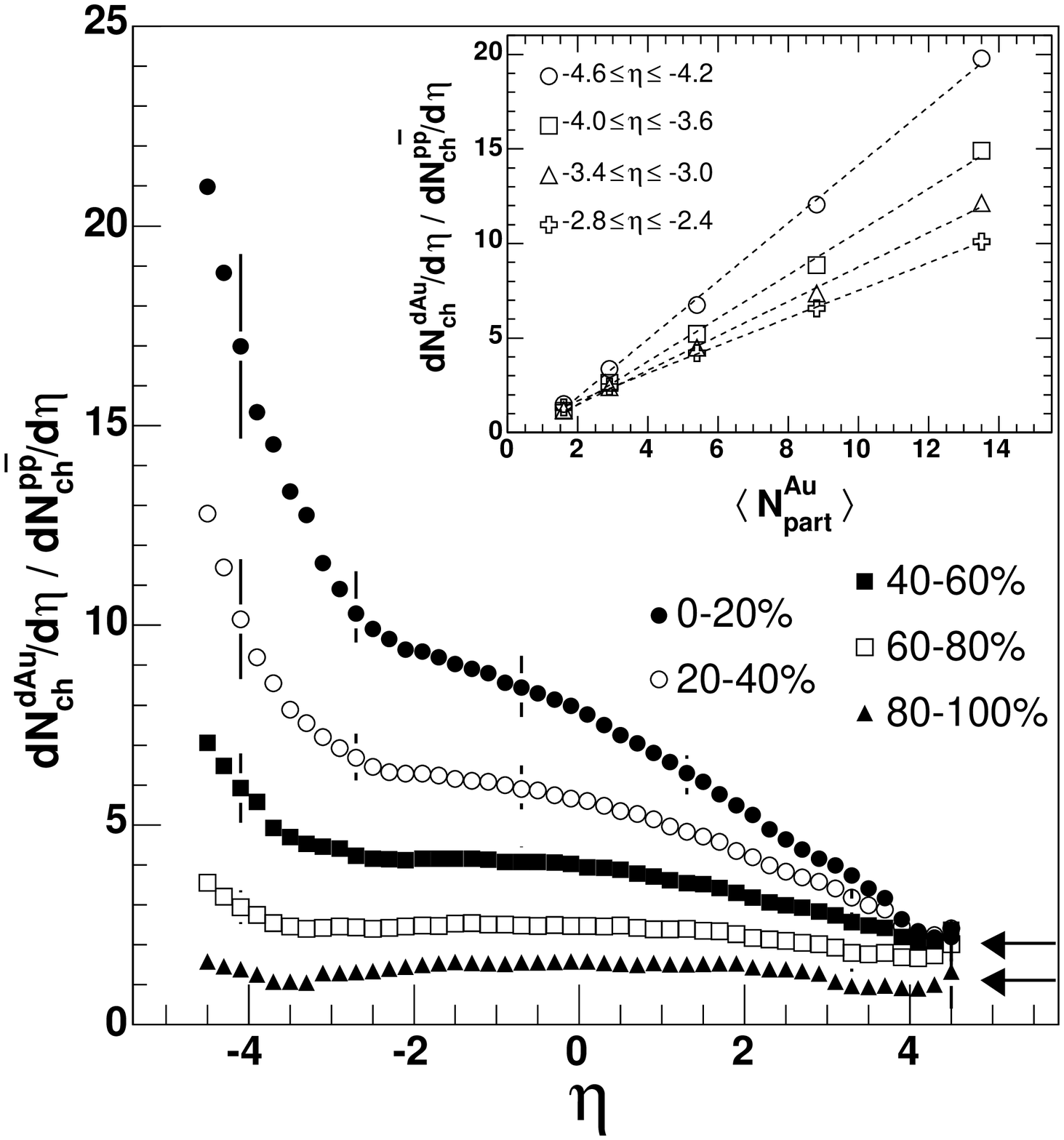}
\caption{
PHOBOS results on d+Au
\label{dAu_cent_Fig4}
}
\end{minipage}
\end{center}
\end{figure}

Other $\jpsi$ data suggests that this is not the case, at
least for $\jpsi$.  PHOBOS data has also shown that the yield
inclusive charged particles ($\nch$) scale approximately linearly
with $\np$ in d+Au and Au+Au, but only when one integrates the
yield over the full phase space\cite{phobos-univ}.  In limited regions, this
linear scaling is clearly broken.  In fact, in the forward region
of d+Au collisions, increasing the number of participants
strongly suppresses the yield in the forward direction. 
This is compensated by an increase in yield in the backward
direction, holding the total yield per participant constant
\cite{Back:2004mr}.  
Similar ``long range''
correlations of particle yields with centrality are also seen 
in PHENIX, where they calculate $R_{CP}$ (the ratio of
yields in central and peripheral events, normalized by the 
number of binary collisions) for $\jpsi$ as well
as stopped hadrons in the muon system, which have $1<p_T<3 GeV$,
as shown in Fig.~\ref{brahms_phenix_rcpJpsi}~\cite{deCassagnac:2004kb}.
It appears again that $\jpsi$ production acts very similarly
to soft particles, at least away from mid-rapidity.  
This compensation of suppression in the forward region with enhancement
in the backward direction may lead to an approximate constancy of 
total $\jpsi$ production with $\np$ or $\nc$ in d+Au collisions.
Taken together with the  various claims
made by Gazdzicki et al~\cite{Gazdzicki:1999rk} and Braun-Munzinger et al
\cite{Andronic:2003zv} who argued that $\jpsi$ production 
actually shows approximate $\np$-scaling, suppression measurements in A+A
collisions may be substantially distorted by limited kinematic acceptance.

If in fact $\np$ scaling is relevant for
$\jpsi$ production in A+A collisions, this may signal
a mechanism better described by statistic models than by parton
model calculations.
Of course, interpreting $\jpsi$ production statistically opens
up the possibility of charm recombination, especially at RHIC
when the charm yields per events are quite large compared with
SPS energies~\cite{Thews:2000rj}.  
The current PHENIX data~\cite{Adler:2003rc} has very large error
bars, and so it is somewhat premature to rule out recombination
scenarios.  The Run 4 data with 10x the statistics should allow
more definitive statements to be made.


\subsection{Open Charm Production}

\begin{figure}[t]
\begin{center}
\begin{minipage}{70mm}
\includegraphics[width=7cm]{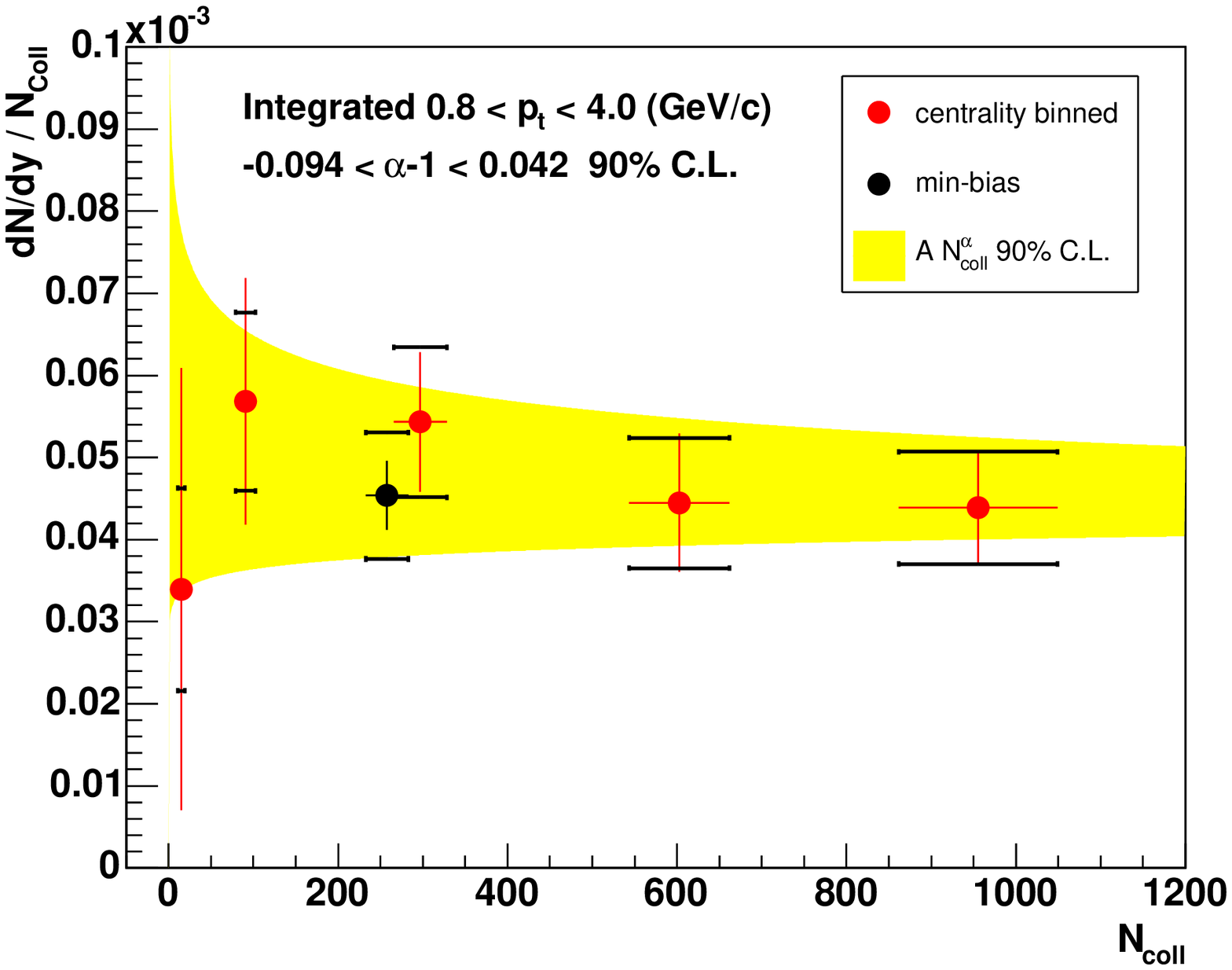}
\caption{
Collision scaling of open charm reported by PHENIX.
\label{auau_dndy_vs_ncoll}
}
\end{minipage}
\hspace{\fill}
\begin{minipage}{70mm}
\includegraphics[width=7cm]{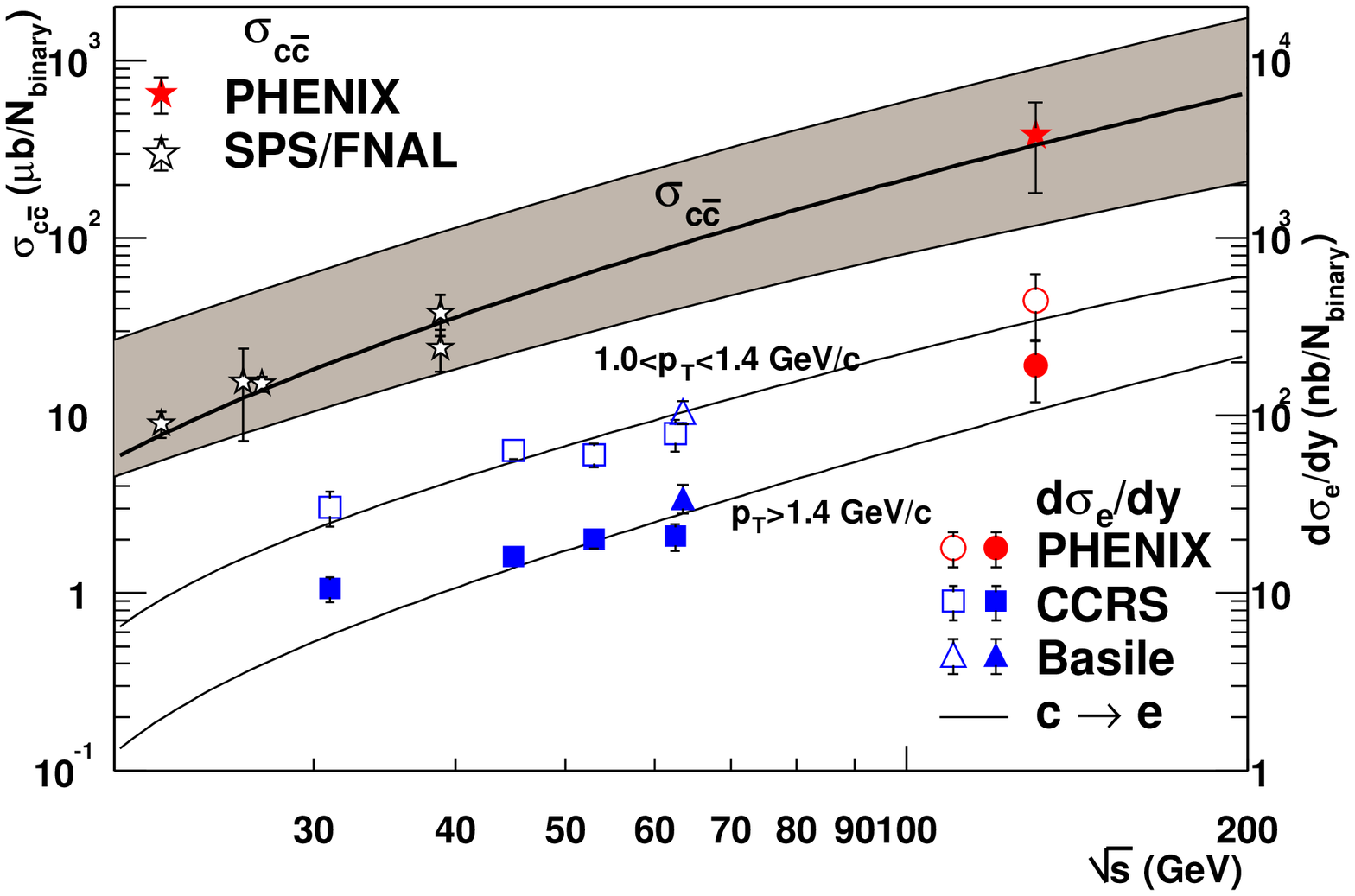}
\caption{
Measurement of the total charm cross section in
Au+Au collisions at $\snn=130$ GeV from PHENIX.
\label{phenixcc-fig4}
}
\end{minipage}
\end{center}
\end{figure}

While one may have expected closed charm production to be sensitive
to final state suppression effects, similar considerations do not
apply to open charm.  Once the quantum numbers have been liberated
in the initial state, they only decay via weak processes on time
scales much longer than the lifetime of the system.
Thus, open charm may well serve the same purpose as Drell-Yan
production and act as a proper reference for the modification
of hard processes, especially if it is found to scale with the
number of binary collisions.  The high-$p_T$ suppression seen
for light hadrons is not expected to show up in the spectrum of
hadrons with heavy quarks (D and B mesons) due to the ``dead cone''
effect discussed above~\cite{Dokshitzer:2001zm}

Open charm has been measured both by PHENIX and STAR, by a variety
of techniques.  PHENIX 
has focused mainly on the extraction of a prompt electron
signal at high $p_T$, which primarily come from charm and beauty decays.
They have performed a measurement of these ``non-photonic'' electrons
out to 5 GeV in the full range of systems offered at RHIC (p+p, d+Au,
Au+Au) and have found that the spectral shape is similar in all of them
\cite{Adcox:2002cg,Averbeck,Adler:2004ta}.
More importantly, they have found that the overall yield seems to
scale linearly with $\nc$, as shown in Fig. \ref{auau_dndy_vs_ncoll}.  
This let them extract a production
cross section per binary collision in heavy ion reactions which is
consistent (within large errors) with pQCD-based extrapolations of
p+p data from Fermilab and the ISR, shown in Fig.~\ref{phenixcc-fig4}.

\begin{figure}[t]
\begin{center}
\begin{minipage}{70mm}
\includegraphics[width=7cm]{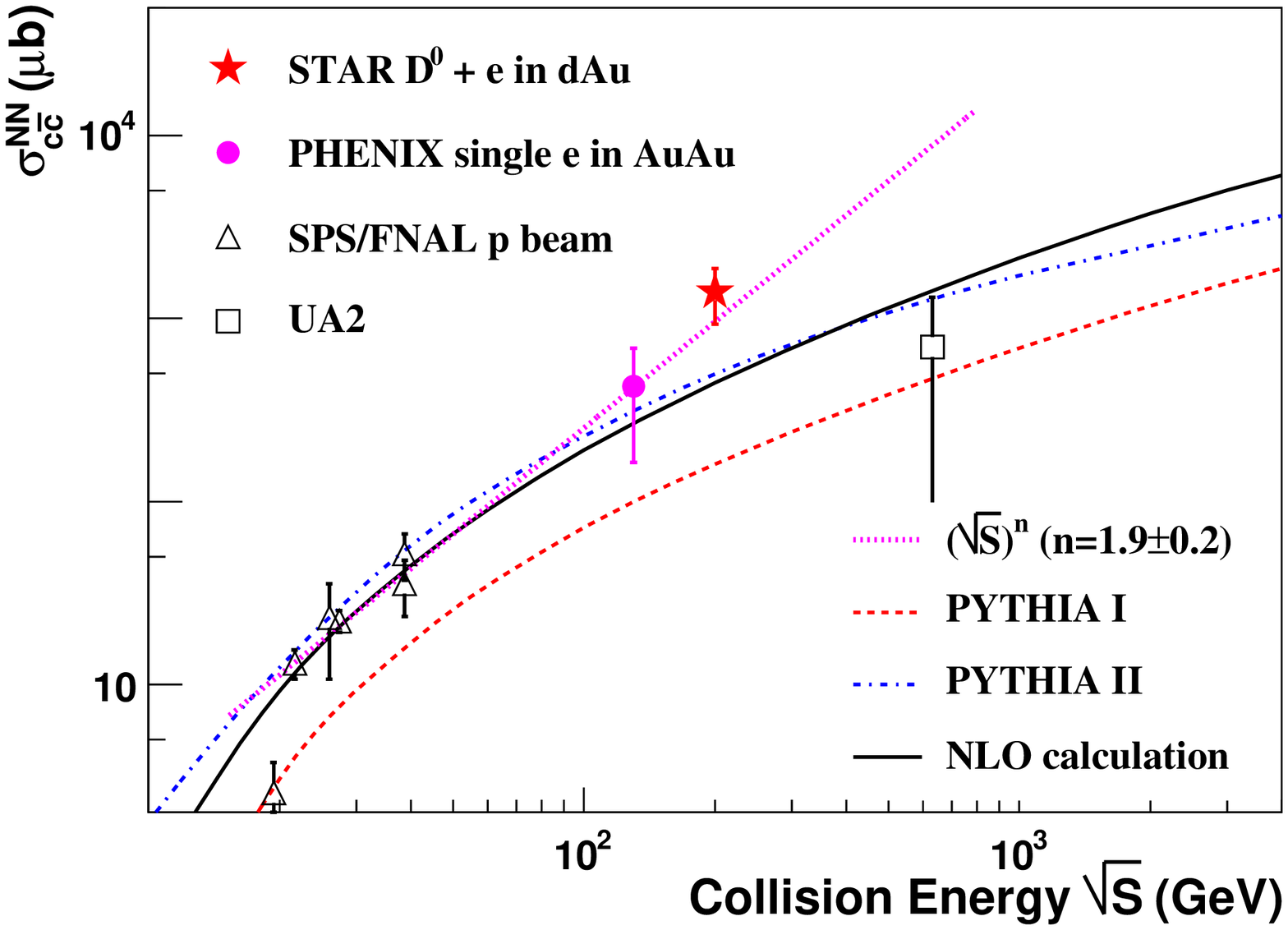}
\caption{STAR measurement of the total charm 
cross section.
\label{Xsec0529}}
\end{minipage}
\hspace{\fill}
\begin{minipage}{70mm}
\includegraphics[width=7cm]{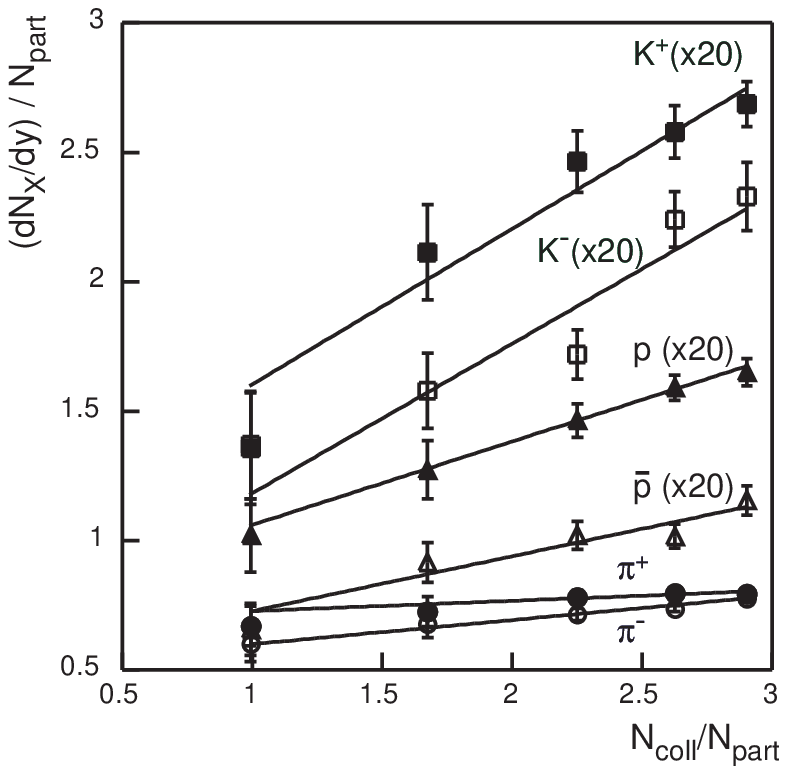}
\caption{
Mid-rapidity yields of various particle species vs. $\nc/\np$
from PHENIX at $\snn=130$ GeV .
\label{dndy-scale-02}
}
\end{minipage}
\end{center}
\end{figure}

STAR has focused both on non-photonic
electron signals as well as direct reconstruction of high-$p_T$ 
D meson decays~\cite{Adams:2004fc,Zhang}. 
By a combination of these methods, they have also
attempted to estimate the total charm production cross section per
binary collision in d+Au collisions, shown in Fig. \ref{Xsec0529}.  
While the STAR and PHENIX
result are consistent with each other within large systematic errors, the STAR
result is substantially higher (by a factor of 2-3) 
than the pQCD extrapolation of
lower-energy data, which describes the PHENIX Au+Au data.  
Still, although this looks like a potential crisis for pQCD, 
one must always keep in mind that charm measurements at new machines
are often quite uncertain as experiments make their first round
of measurements, leading to a large variance between them (e.g.
at the ISR, as we saw in Fig.~\ref{vogt-ccbb}).  Confirming measurements over a variety of systems
and final state charmed particles may be necessary before
the charm cross section at RHIC energies can be established to 
a high precision.

\section{Strangeness and Hard Scaling}

In the previous sections, we saw that particles with hidden charm 
(e.g. $\jpsi$) scale in
some ways like $\np$ while,
open charm production rates seem to
show scaling with $\nc$.  Prima facie, this is proof of
``hard'' scaling (presumably due to the ``dead cone'' effect)
and thus charm seems to serve well as a reference
for the suppression of other processes.  Does this suggest that
charm can truly serve as a calibrated reference?
It will be argued in this section that the situation with 
the production of non-light flavors is somewhat more subtle
than it would first appear.

The first observation related to this issue is that charm is not the
only process that scales much faster than the number of participants.
While strangeness production is not considered a ``hard'' process,
in that the relevant mass scale is on the order of the QCD scale,
it has been observed that the number of kaons per participant
increases with the number of participants, almost by a factor of two
between peripheral and central events, e.g. as shown by PHENIX
at $\snn=130$ GeV in Fig.~\ref{dndy-scale-02}~\cite{Adcox:2003nr}.

However, pure binary collision scaling would imply an increase
per participant of a factor of 6 relative to p+p collisions, 
something which is typically not seen for strange particles.
This begs the question of what kind of scaling is in fact 
observed, and whether it bears a simple relationship to the
initial-state nuclear geometry.

\begin{figure}[t]
\begin{center}
\includegraphics[width=12cm]{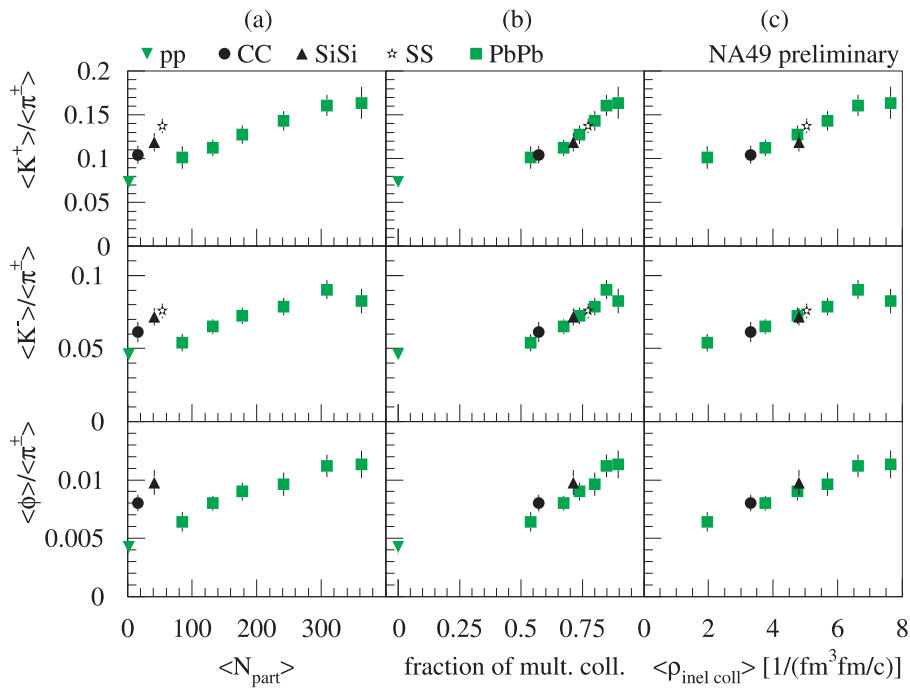}
\caption{
Data on $K^{+}/\pi^{\pm}$, $K^{-}/\pi^{\pm}$,
and $\phi/\pi^{\pm}$ vs. system size in NA49
at Pb+Pb collisions at 158 AGeV.
\label{ratios_sys_news1}
}
\end{center}
\end{figure}

\begin{figure}[t]
\begin{center}

\begin{minipage}{70mm}
\includegraphics[width=7cm]{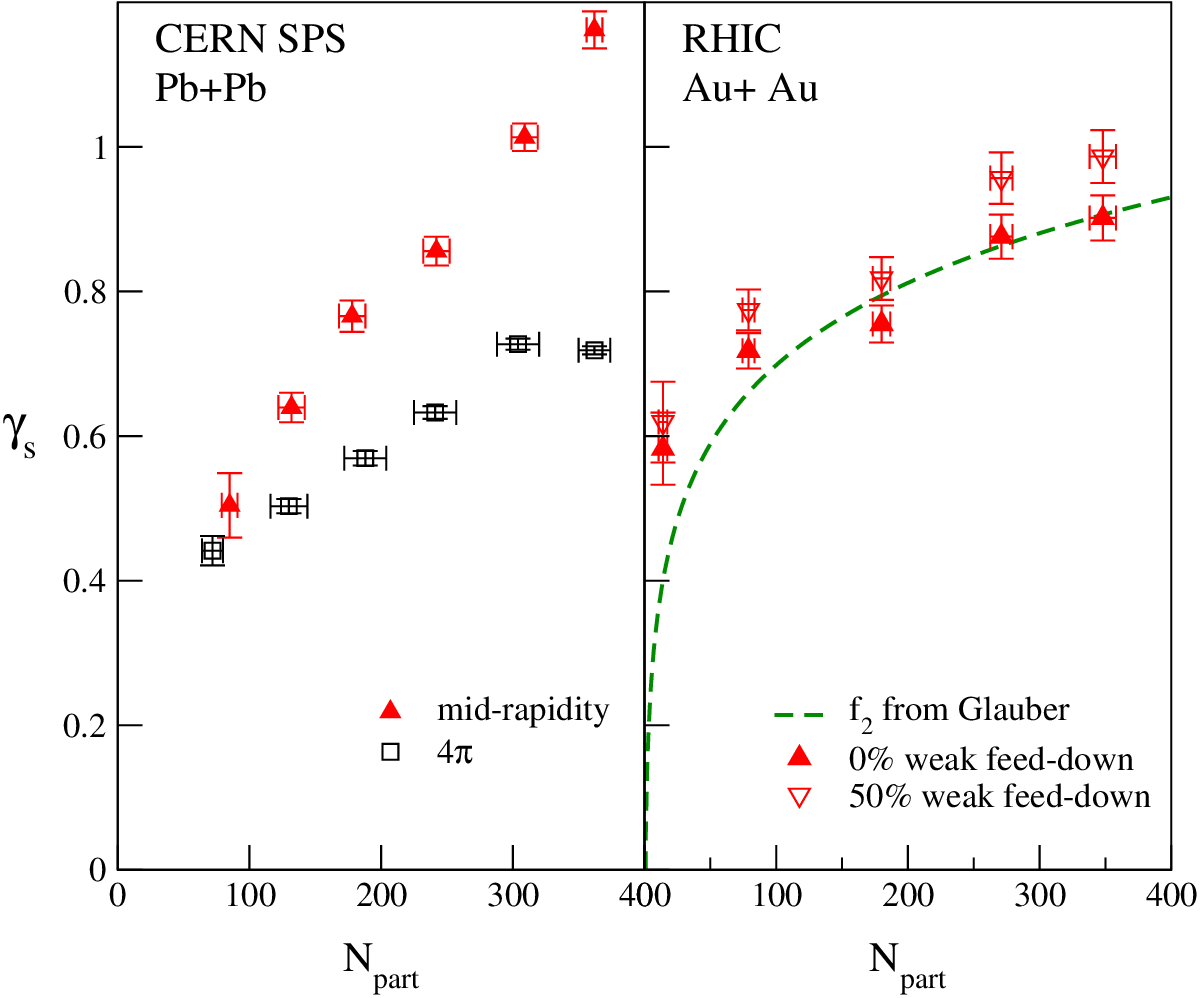}
\caption{
Centrality dependence of $\gamma_s$ at SPS
and RHIC energies.
\label{gammas_mod}
}
\end{minipage}
\hspace{\fill}
\begin{minipage}{80mm}
\includegraphics[width=7cm]{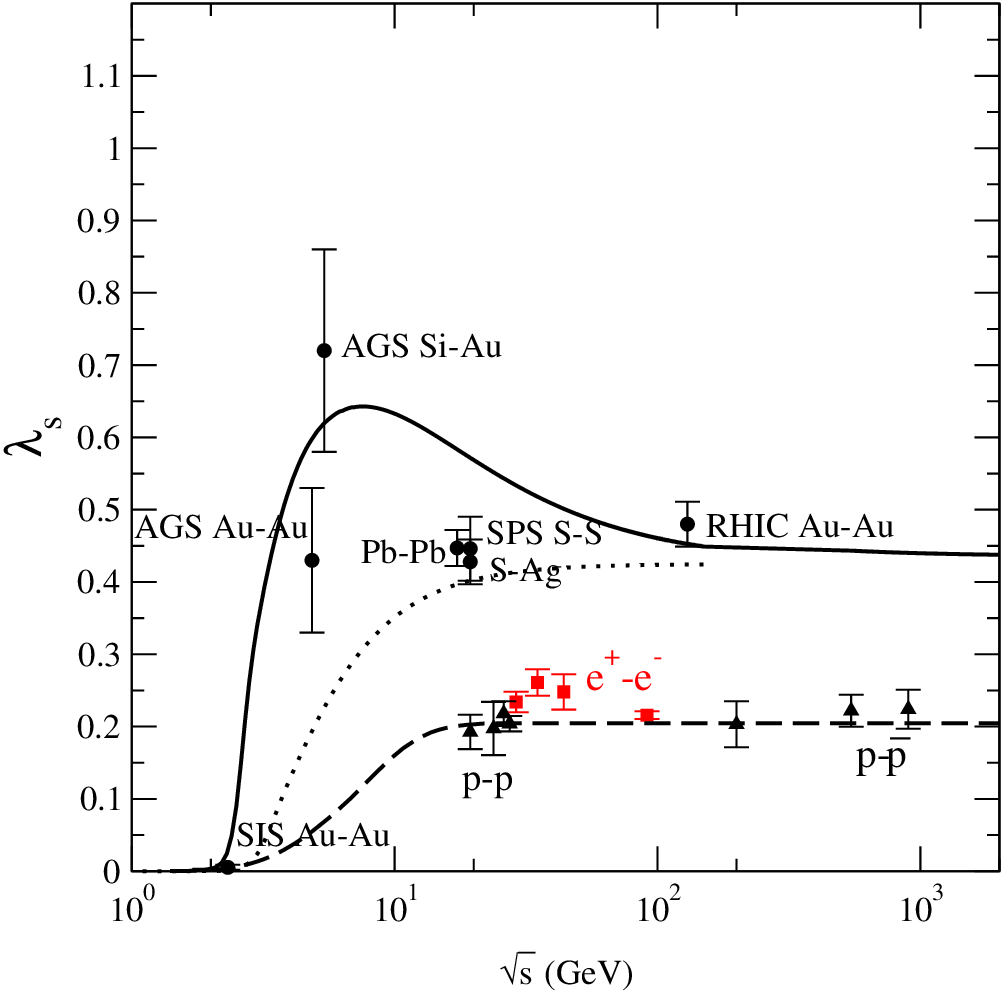}
\caption{
Compilation of $\lambda_s$ as a function of 
beam energy for $\pp$, $\epem$ and A+A
reactions.
\label{jaipur2_fig4}
}
\end{minipage}
\end{center}
\end{figure}

One hint about the relevant control variables was given by an
analysis of strangeness production over a large range of system
sizes by the NA49 collaboration, shown in Fig.~\ref{ratios_sys_news1}
\cite{Hohne:2002pb,Alt:2004wc}.  
They studied the $\kpi$ ratio
as a function of various centrality variables and found that
$\np$ is not a proper scaling variable to connect smaller
systems like $Si+Si$ and $C+C$ to $Pb+Pb$.  Two variables which
do work are the fraction of multiply-struck participants (which
we call $f_2$), and the space-time collision density, as shown
in the second and third columns of Fig.~\ref{ratios_sys_news1}.  
While the data is equally consistent with both 
of these variables, it should be pointed out that they have 
slightly different physical pictures.  The scaling with $f_2$ suggests
that strangeness enhancement (or the lifting of strangeness
suppression) is a purely geometrical effect, essentially 
independent of the particle density.  The collision density
is more related to the number of times nucleons scatter
on their way through the oncoming nucleus, but it requires more
assumptions (as expressed through a transport code) to 
derive the scaling quantity.  For this reason, $f_2$ may be
the more relevant variable, as it requires the fewest additional
physics assumptions to achieve a reasonable scaling behavior.
Intriguingly, centrality-dependent thermal fits have found that $\gamma_s$ is
quantitatively very close to $f_2$ over a large range of centralities~\cite{Cleymans:2002hy},
as shown in Fig. \ref{gammas_mod}.  This suggests that $\gamma_s$ should be
less than unity (by at least $10\%$) for all
observable heavy ion collisions, if only due to the ``skin'' of singly-struck
nucleons.

And yet, the observation of a scaling behavior does not explain
the mechanism for why it works.  It is beyond the scope of this
work to propose a detailed model for the success of $f_2$-scaling,
but it is useful to imagine what happens to a nucleon in the context
of a heavy ion collision.  The first collision certainly excites 
the nucleon into a massive state that decays into hadrons.  This
scenario is the essence of the wounded nucleon model of particle production.  
The second collision, however, involves an interaction of this
excited nucleon with another nucleon (also possibly excited). 
While the original nucleon should have a well-understood parton
structure (as explored via DIS), the multiply struck nucleon
is not something about which we have direct information.  It
is possible that this object will have a severely distorted 
valence structure.

Some insight may have already been provided 
by an analysis of the strange sea of the
nucleon from NuTeV~\cite{Goncharov:2001qe}.  By a study of charm 
production in $\nu+Fe$ interactions, they extracted the
{\it strange} content of the nucleon (via the flavor changing
charged current $\nu+s \rightarrow \mu + c$).  The overall
fraction of the sea occupied by strange quarks was quantified
by a scale factor $\kappa = (2s)/(u+d)$, a quantity identical
to the Wroblewski factor $\lambda_s$.  The measurements showed
values of $\kappa = 0.36\pm0.05$ in $\nu+N$ interactions and
$\kappa = 0.38\pm0.04$ in $\overline{\nu}+N$.  When one compares
this to a broad compilation of $\lambda_s$ extracted from A+A,
p+p and $\epem$ data, it is observed that the strangeness
content of the sea is intriguingly close to the asymptotic
value seen in A+A collisions, which is about 0.44, as shown in
Fig.~\ref{jaipur2_fig4}~\cite{Cleymans:2002mp}.

\section{Outlook and Conclusions}

In conclusion, we have reviewed various aspects of heavy flavor production
in heavy ion collisions in the context of elementary collisions.
The interplay between theory in experiment in $\epem$ and $\pbarp$ data has
been discussed and it is found that the pQCD description is not straightforward
even now, although substantial progress is being made with NLO
calculations.
Quarkonium suppression in p+A and A+A show some very puzzling features, if the
primordial charm quarks are truly produced by factorisable hard processes.
It is mysterious how $\jpsi$ production shares features with ``soft''
inclusive particle production.
It is equally mysterious how open strangeness also scales much faster
than $\np$, an effect which should not be entirely ignored when considering
the $\nc$-scaling of open charm.
It is evident that heavy ions, by offering an environment where nucleons are
struck multiple times (unlike elementary collisions), provide a unique 
perspective on the production of new flavor quantum numbers over a wide
range of momentum scales.  So while the present occasionally seems ``strange'',
although often ``charming'', the future of heavy flavor studies at RHIC
certainly looks ``beautiful'', with upgrades
and planned searches for the $\Upsilon$ and leptons from $B$ decay
in the large statistics data sets
taken in RHIC Run 4\cite{Calderon}.

\section{Acknowledgments}
The author would like to thank the organizers, Federico Antinori,
Steffen Bass, Rene Bellwied, Thomas Ullrich, Julia Velkovska and
Urs Wiedemann, who invited
a non-expert sight-unseen to give a non-traditional talk on this subject.
Invaluable help was given by David Silvermyr and Raphael de Cassagnac,
who supplied me with preliminary PHENIX results. Special thanks
to 
Ralf Averbeck, 
Terry Awes,
Mark Baker, 
Jean Cleymans,
and Paul Stankus for discussions related to this work.


\end{document}